\documentclass[prd,aps,preprint,amsmath,nofootinbib,amssymb,eqsecnum,showkeys,tightenlines]{revtex4-1}
\usepackage{array}
\usepackage{amsmath, latexsym, amssymb, hyperref, graphicx, color, multirow, makecell, diagbox}
\usepackage{array}
\usepackage{tabularx}
\usepackage[T1]{fontenc}
\usepackage[capitalize]{cleveref}
\usepackage{booktabs}
\usepackage{multirow}
\usepackage{graphicx}
\usepackage{subcaption}
\usepackage{siunitx} % 用于对齐数字小数点
\usepackage{multirow}
\usepackage{slashed}
\numberwithin{equation}{section}
\usepackage{verbatim}
\allowdisplaybreaks

\def\nnb{\nonumber}

\newcommand{\be}{\begin{equation}}
\newcommand{\ee}{\end{equation}}
\newcommand{\bea}{\begin{eqnarray}}
\newcommand{\eea}{\end{eqnarray}}
\newcommand{\ba}{\begin{array}}
\newcommand{\ea}{\end{array}}

\def\be{\begin{equation}}
\def\ee{\end{equation}}
\def\bea{\begin{eqnarray}}
\def\eea{\end{eqnarray}}
\def\nnb{\nonumber}
\def\xslash\sharp1{{\rlap{$\sharp1$}/}}

\begin{document}
\title{Comments on "Unified neutrino mixing and approximate $\mu-\tau$ reflection symmetry"
(Mod. Phys. Lett. A, 40 (2025) 26, 2550097
[arXiv:2502.18029 [hep-ph]]) }
\author{Chao-Shang Huang}
\email{csh@itp.ac.cn}
\affiliation{
\normalsize CAS Key Laboratory of Theoretical Physics, Institute of Theoretical Physics, \\
 Chinese Academy of Sciences, Beijing 100190, China}
\author{Wen-Jun Li}
\email{liwj24@163.com}
\affiliation{
\normalsize Physics school, Henan Normal University, Xinxiang 453007, China}

\begin{center}

\begin{abstract}
Resolving mass ordering is an important issue in the neutrino physics. In Ref.\cite{yt} the authors
investigate the phenomenology of a unified neutrino mixing framework and reveal that the predicted
sum of neutrino masses derived from an approximate $\mu-\tau$ reflection symmetric flavor neutrino mass
matrix based on the unified neutrino mixing with an inverted mass ordering, is excluded
from DESI2024 and Supernova Ia luminosity distance data. We note that in Ref.\cite{yt} an error is present in Eq.(20), {\it  i.e.}, the expression for $ M_{23}^{\mu-\tau} $, a similar error also appears in Eq.(26) for $M_{13}^{\mu-\tau}$.
That is, the condition that $M_{ee}$ and $M_{\mu\tau}$ are real is omitted. We impose this condition and analyze its consequences. Using the newest data\cite{azz}, it is obtained that the predicted sm derived from the approximate $\mu-\tau$ reflection symmetric neutrino mass matrices $M_{23}$ for both NO and IO are allowed. We note also that their conclusion is invalid, as it is based on outdated data.

\end{abstract}
\end{center}

\maketitle

\section{introduction}\label{sec1}
In recent years much progress has been made in experiments on neutrino oscillations. However, to date, in addition to the absolute neutrino mass scale and two Majorana cp-violating phases, neutrino mass ordering is also not determined by experiments on neutrino oscillations. Consequently, theoretically resolving the neutrino mass ordering within phenomenological models is of significant value for guiding and informing future experimental efforts.
 In the literature, some papers\cite{phemo,hl,hcs} discuss the neutrino mass ordering. For example, Huang and Li\cite{hl} investigate the constraints to model parameters from the neutrino mass hierarchy. Huang\cite{hcs} computes the effective Majorana neutrino mass in NO (normal order of neutrino masses: $m_1<m_2<m_3$) and IO (inverted order of neutrino masses: $m_3< m_1< m_2$) both cases. In Ref.\cite{yt} the authors
investigate the phenomenology of a unified neutrino mixing framework with an approximate $\mu-\tau$ reflection symmetric flavor neutrino mass matrix and obtain a conclusion: an inverted mass ordering is excluded from DESI2024 and Supernova Ia luminosity distance data. We note that in Ref.\cite{yt} an error is present in Eq.(20), {\it i.e.}, the expression for $ M_{23}^{\mu-\tau} $, a similar error also appears in Eq.(26).
% there is an error in $M_{23}^{\mu-\tau}$, (20) (similar to (26)) in Ref.\cite{yt}.
That is, the condition that $M_{ee}$ and $M_{\mu\tau}$ are real is not added. We add the condition and analyze its consequences. We note also that their conclusion is invalid, as it is based on outdated data.

In this paper we review the ${\mu-\tau}$ reflection symmetry and its limitations to neutrino mass matrix, and give our comments on Ref.\cite{yt}.

In Section ~\ref{sec2}, we review the ${\mu-\tau}$ reflection symmetry and its limitations. Consequences of real $M_{ee}$ and $M_{\mu\tau}$ are given in Section ~\ref{sec3}. In Section ~\ref{sec4} we calculate the effective Majorana neutrino mass and give our comments on Ref.\cite{yt}. Finally we present
summary and outlook in Section \ref{sec5}.

\section{The $\mu-\tau$ reflect symmetry}\label{sec2}
It is well known that $|U_{\mu i}|=|U_{\tau i}|$
(for i=1,2,3) known as the $\mu-\tau$ reflection (or interchange ) symmetry\,\cite{hcs,bmv,ma}. Babu, Ma, and Valle (BMV) \cite{bmv,ma} suggest a Majorana mass matrix as
\bea
M_{\nu}=\left(\begin{array}{ccc}
a & r & r^* \\
r & s & b \\
r^* & b & s^*
\end{array}\right),
\eea
where $r$ and $s$ are in general complex while $a$ and $b$ remain real. It is proved that the mass matrix of BMV always yields the $\mu-\tau$ reflection symmetry.

In Ref.\cite{yt}, $M_{23}^{\mu-\tau}$ satisfies the exact $\mu-\tau$ reflection symmetry. For the expression of $M_{23}^{\mu-\tau}$, see Eq. (20) in Ref.\cite{yt}. Therefore, $M_{ee}=a$ and $M_{\mu\tau}=b$ should be real. However, the authors overlook the necessary condition. In this section we solve the equations $a_1=Im[a]=0$ and $a_2=Im[b]=0$.

 From Eq.(20,22) in Ref.\cite{yt}
\bea
 a &=& M_{ee}=\frac{ nmee}{dmee}~,  \\
nmee &=& a^2 m_1 + 2 m_2 \,ca \,cs1^2+ 2 m_3 (cb \,sn1^2 )  +I(\, 2 m_2  \,sa \,cs1^2 + \, 2 m_3\,sb  \,sn1^2)~,\nnb \\
dmee &=& (a^2+2)~, \nnb \\
\label{mee}
b &=& M_{\mu\tau}= M_{ee} - (a -\frac{1}{a}) M_{e\mu} +\frac{1}{a}M^*_{e\mu} - M_{\mu\mu}=\frac{nmem}{dmem}~,  \\
nmem &=& 2 a m_1 - m_2\biggl[a (cs1^2 - sn1^2) + a + I \sqrt{a^2 + 2} 2 cs1 \,sn1\biggl](ca+I sa)  \nnb \\
&& - m_3\biggl[-a (cs1^2 - sn1^2) + a -I \sqrt{a^2 + 2}  2 cs1 \,sn1\biggl](cb+I sb)~, \nnb \\
dmem &=&2 (a^2 + 2)~, \nnb \label{mem}
\eea
where $ ca = \cos 2\alpha, sa = \sin 2\alpha, cb=\cos 2\beta, sb=\sin  2\beta,
cs1 =\cos \theta,  sn1 = \sin \theta, \theta$ and $a$ are parameters in the mixing matrix.  $\alpha,\beta$ are Majorana phases.

so
\bea a_1 &=& Im[a]=\frac{2 cs1^2 m_2\, sa + 2 m_3\,  sb \,sn1^2}{2 + a^2}~,\label{a11} \\
a_2 &=& Im[b]=\frac{a^2 (m_2\,  sa +  m_3 \, sb) + 2 cs1^2 m_3\,  sb + 2 m_2\,  sa\, sn1^2}{4 + 2 a^2}. \label{a21} \eea

\section{Consequences of real $a$ and $b$}\label{sec3}
Solving Eqs.~(\ref{a11},\ref{a21}), it obtains that
\bea
sa &=& \frac{m_3\, sb \,sn1^2}{m_2\,cs1^2 }, \label{xy}\\ sb &=& -\frac{m_2 \,sa (a^2 + 2 sn1^2)}{m_3(a^2 + 2 cs1^2) }.\label{sab}\eea
Eq.~(\ref{xy})-(\ref{sab}) leads to
\bea &&cs1^2 = sn1^2,\nnb
\\&&sb = -\frac{ m_2 \,sa}{m_3}.\label{sb}\eea
That is, the parameter $\theta$ in the mixing matrix is determined to be $\pm\frac{n\pi}{4}(n=1,2....),$ and $sb$ with undetermined $sa$.

We now compute the absolute neutrino mass scale constrained by cosmological observations. We define the sum of neutrino masses, $sm=\sum m_i=m_1+m_2+m_3$. For NO,$(m_1, m_2, m_3) = (m_1, \sqrt{\delta m^2_{21} + m_1^2} ,\sqrt{\delta m^2_{31} + m_1^2}) $. For IO, $(m_1, m_2, m_3)=(m_1,$\,\,\,\\$ \sqrt{\delta m^2_{21} + m_1^2},\sqrt{ \delta m^2_{32} + m_2^2})$. The upper limit of $\sum m_i$, 0.09 eV, from\cite{valpal}, and 0.12 eV, from\cite{agh}. Recently, in Ref.\cite{desi}, from the combination of DESI and CMB, it is derived that $95\%$ upper limits on the sum of neutrino masses, is smaller than 0.064 eV, \, assuming $\Lambda$CDM and $\sum m_i$ is smaller than 0.16 eV in the $\omega_0 \omega_a$ model.
% We do not consider the case since it depends on cosmology models.
 For $\sum m_i=0.12$ eV \cite{agh}, we obtain $ (m_1,m_2,m_3)= (0.03021,0.03142, 0.05837)$ for NO and $ (m_1,m_2,m_3)= (0.05197, 0.05268, 0.01535)$ for IO. The upper limit of $\sum m_i= 0.09 eV$\,\cite{valpal}, leads to $ (m_1,m_2,m_3)$\\$= (0.01752, 0.01955,0.05293)$ for NO and no solusion for IO.  For $\sum m_i=0.064$ eV \cite{desi}, $ (m_1,m_2,m_3)=( 0.004232,0.009639,0.05013)$ for NO and also no solution for IO.

 Have known neutrino masses and the Majorana CP-violating phase in Eq.~(\ref{sb}), we chose some values for sa and compute sb, the results are list in Table I.

\section{The effective Majorana neutrino mass and the sum of the neutrino masses}\label{sec4}

Have known the Majorana CP-violating phases $\alpha, \beta$ and neutrino masses ($m_1, m_2, m_3$),  the effective Majorana neutrino masses $M_{ee}$ can be directly computed. 
Its absolute value is experimentally accessible\footnote{It is also referred to as the effective neutrino mass of the neutrinoless double $\beta$ decay.}  via neutrinoless double-beta decay $(0\nu\beta\beta)$ searches.
%($0\nu 2\beta$) experiments
\be \label{mee}
M_{ee} = m_1 c^2_{12} c^2_{13} e^{i 2\alpha} + m_2 s^2_{12} c^2_{13} e^{i 2\beta} + m_3 s^2_{13} e^{-2 i\delta}. \\
\ee
To input experimental data in Eq.~(\ref{mee}),  we could obtain $M_{ee}$  which is shown in Table II.

\begin{table}[htbp]
\centering
\caption{The values of $sa(=\sin 2\alpha)$ and $sb(=\sin 2\beta)$ with $\sum m_i =(0.064,0.09,0.12) ,$ where $m_i$ is $m_i/eV, i=1,2,3.$}
\label{tab:values}
\begin{tabular}{c| c | w{c}{2cm} | w{c}{2cm} | w{c}{2cm} |}
\hline
\toprule
\multirow{2}{*}[-0.5ex]{} & \multirow{2}{*}[-0.5ex]{Parameter} & \multicolumn{3}{c|}{$\sum m_i (eV)$} \\ \cline{3-5}
& & \textbf{0.064} &\textbf{ 0.09} &\textbf{ 0.12} \\ \hline
\multirow{6}{*}{NO} & $m_1(eV)$ & \textbf{0.004232 }& \textbf{0.01752 }&\textbf{ 0.03021} \\ \cline{2-5}
& $m_2(eV)$ &\textbf{ 0.009639 }&\textbf{ 0.01954} &\textbf{ 0.03142} \\ \cline{2-5}
& $m_3(eV)$ &\textbf{ 0.05013 }& \textbf{0.05293} &\textbf{ 0.05837} \\ \cline{2-5}
& $sa$ & \textbf{0.3420} & \textbf{0.3420} & \textbf{0.3420} \\ \cline{2-5}
& $sb$ & \textbf{-0.06577} & \textbf{-0.1263} & \textbf{-0.1841} \\  \hline
\multirow{6}{*}{IO}& $m_1(eV)$ & ---  & --- & \textbf{0.05197 }\\ \cline{2-5}
& $m_2(eV)$ & ---  & ---  & \textbf{0.05268} \\ \cline{2-5}
& $m_3(eV)$ & ---  & ---  & \textbf{0.01535} \\ \cline{2-5}
& $sa$ & ---  & --- & \textbf{0.1736} \\ \cline{2-5}
& $sb$ &  ---  & ---  & \textbf{-0.5960 }\\  \hline
\end{tabular}
\end{table}

\begin{table}[htbp]
\centering
\caption{The effective Majorana neutrino mass $M_{ee}$ with $\sum m_i =(0.064,0.09,0.12) $, where $M_{ee}$ is $M_{ee}/(0.01eV)$  and $m_i$ is $m_i/eV , i=1, 2, 3 $.}
\begin{tabular}{c| c| w{c}{3cm}| w{c}{3cm}| w{c}{3cm}|}
\toprule
 \multirow{2}{*}[-0.5ex]{} & \multirow{2}{*}[-0.5ex]{Parameter} & \multicolumn{3}{c|}{$\sum m_i (eV)$} \\ \cline{3-5}
 & & \textbf{0.064} & \textbf{0.09} & \textbf{0.12} \\\hline
 \multirow{6}{*}{NO} & $m_1(eV)$ & \textbf{0.004232} & \textbf{0.01752} & \textbf{0.03021} \\\cline{2-5}
 & $m_2(eV)$ & \textbf{0.009639} & \textbf{0.01954} & \textbf{0.03142} \\\cline{2-5}
 & $m_3(eV)$ & \textbf{0.05013} & \textbf{0.05293} & \textbf{0.05837} \\\cline{2-5}
 & \makecell{$M_{ee}$\\(0.01eV)} & \textbf{0.5939-i0.02772} & \textbf{1.735+i0.2179} & \textbf{2.8909+i0.3990} \\\cline{2-5}
 & \makecell{$|M_{ee}|$\\(0.01eV)} & \textbf{0.5945} & \textbf{1.748} & \textbf{2.918} \\
\midrule
 \multirow{6}{*}{IO} & $m_1(eV)$ & --- & --- & \textbf{0.05197} \\\cline{2-5}
 & $m_2(eV)$ & --- & --- & \textbf{0.05268} \\\cline{2-5}
 & $m_3(eV)$ & --- & --- & \textbf{0.01535} \\\cline{2-5}
 & \makecell{$M_{ee}$\\(0.01eV)} & --- & --- & \textbf{4.701-i3.441} \\\cline{2-5}
 & \makecell{$|M_{ee}|$\\(0.01eV)} & --- & --- & \textbf{4.714} \\
\bottomrule
\end{tabular}
\end{table}

In Ref.\cite{yt}, the approximate ${\mu-\tau}$ reflection symmetric flavor neutrino mass matrix is given in  Eq.(19):
\be M_{23}=
M_{23}^{\mu-\tau} + \Delta M_{23}, \label{m23}  \ee
where $\Delta M_{23}$ is a matrix representing the deviation from the exact ${\mu-\tau}$ reflection symmetry and is shown in Eq.(21) in  Ref.\cite{yt}. The dimensionless parameters $\epsilon_{23}$ is introduced to characterize the violations of the ${\mu-\tau}$ reflection symmetry and if  $0<\epsilon_{23}\leq 0.1$, the authors deem that the flavor neutrino mass matrix $M_{23}$ is satisfied with an approximate ${\mu-\tau}$ reflection symmetry.

To use our corrected results Eq.(\ref{sb}),
we obtain :
\bea
\epsilon_{23} &=& \frac{ ft1}{doc1 }+\frac{ ft2}{doc2},\\
doc1 &=& (2 + a^2) (m_2 \,ca- m_3\,cb)^2 + \biggl[a (2 m_1 -m_2 \,ca-  m_3\,cb)+ 2 \sqrt{2 + a^2} m_2 \,sa \biggl]^2,\nnb\\
doc2 &=& a^2 (2 + a^2) (m_2 \,ca- m_3\,cb)^2 + \biggl(-2 m_1 +  m_2 \,cb+  m_3\,cb+  2 a \sqrt{2 + a^2} m_2\, sa\biggl)^2,\nnb\\
ft1 &=& rft1 + I ift1, \\
rft1  &=& 4 m_2 \,sa \biggl[a \sqrt{2 + a^2} (-2 m_1 +  m_2 \,ca+ m_3\,cb) -2 m_2 \,sa(2+ a^2)\biggl],\nnb\\
ift1 &=& 4 m_2  \,sa  (2 + a^2)(m_2 \,ca- m_3\,cb).\nnb \\ ft2 & = & rft2 + I ift2,\\
rft2 &=& 4 a m_2 \,sa\biggl[-\sqrt{2 + a^2}(-2 m_1 +  m_2 \,ca+ m_3\,cb)-2 a m_2\, sa(2+ a^2)\biggl],\nnb\\
ift2 &=&
 4 a^2 (2 + a^2)  (m_2 \,ca- m_3\,cb) m_2\,sa. \nnb
 \eea

\begin{figure}[htbp]
    \centering
    \begin{subfigure}[b]{0.48\textwidth}
        \centering
        \includegraphics[width=\textwidth]{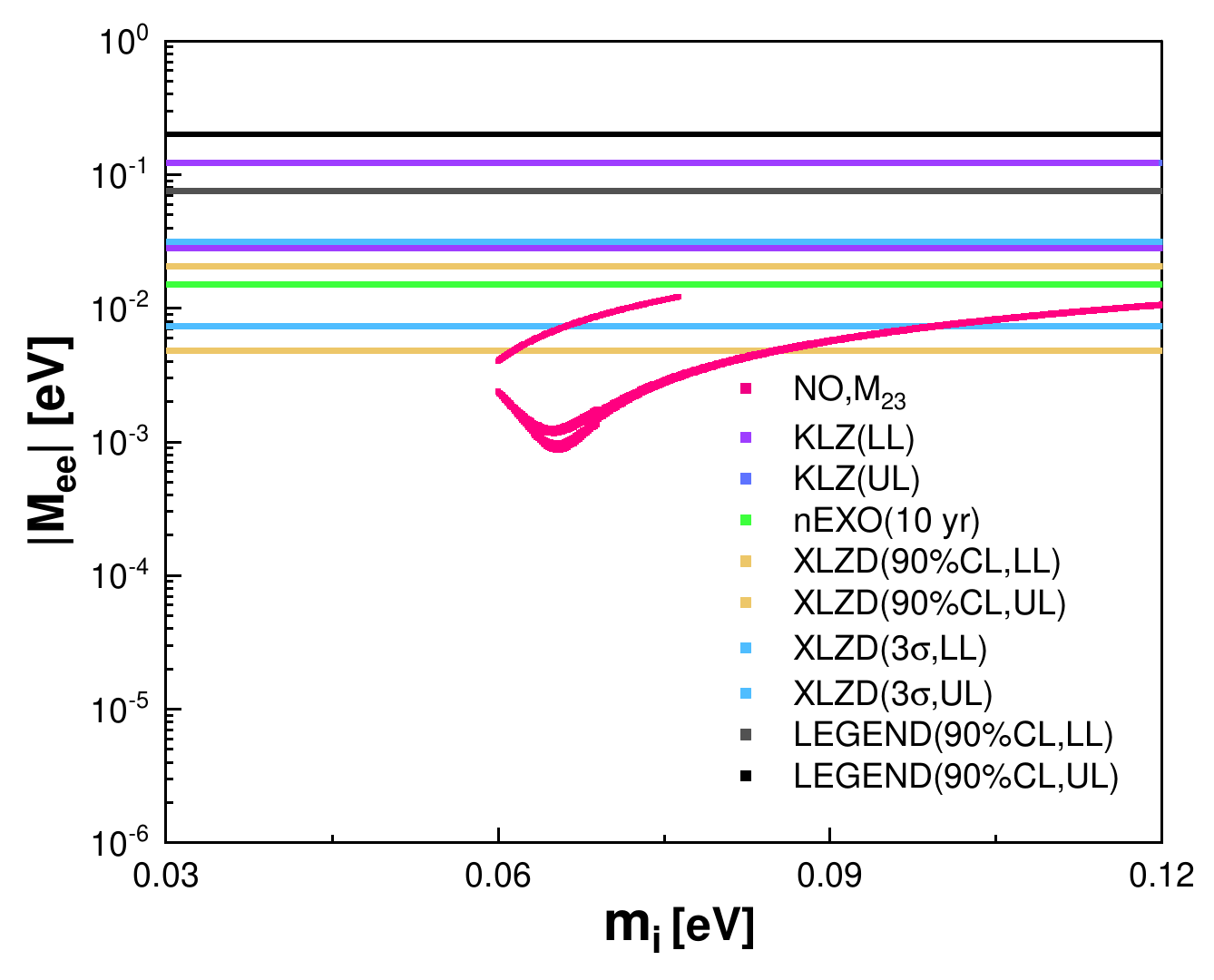}
       % \caption{}
        %\label{fig:suba}
    \end{subfigure}
    \hfill
    \begin{subfigure}[b]{0.48\textwidth}
        \centering
        \includegraphics[width=\textwidth]{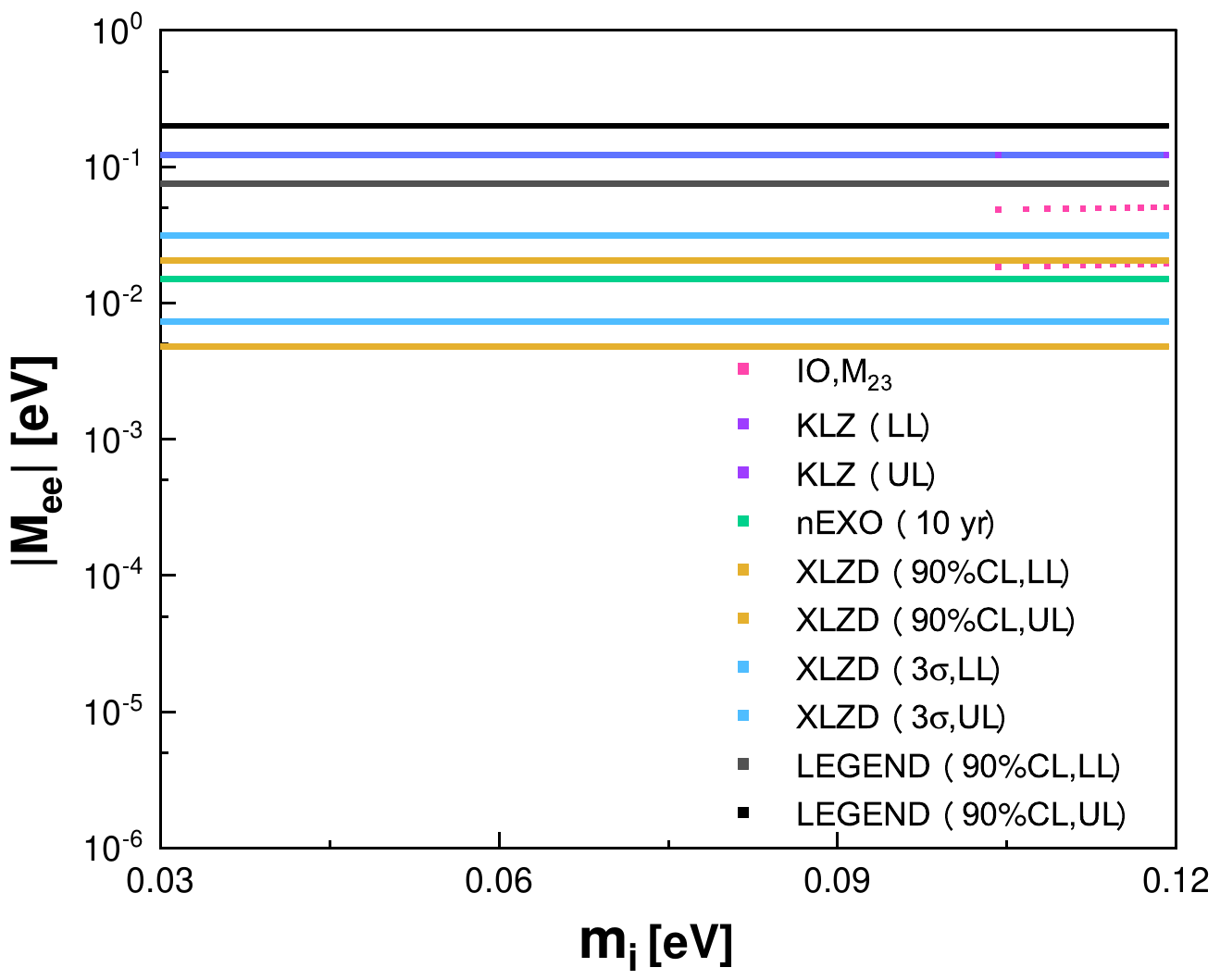}
        %\caption{第二个子图的说明}
        %\label{fig:subb}
    \end{subfigure}
    % 总标题，引用时可用 \ref{fig:overall}
    \caption{Relationship between the effective neutrino mass of the neutrinoless double $\beta$ decay $|M_{ee}|$
and the sum of the neutrino masses
$\sum m_i$ in $M_{23}$ for the NO (left panel) and IO (right panel),
where  the condition of $0<\epsilon_{23}\leq  0.1$ needs to be met.}
    \label{fig:overall}
\end{figure}

The relationship between the effective neutrino mass for neutrinoless double-$\beta$ decay, $|M_{ee}|$, and the predicted sum of the neutrino masses, $\sum m_i$ in the $M_{23}$ parameter space, derived from an approximate $\mu-\tau$ reflection-symmetric flavor neutrino mass matrix within a unified neutrino mixing framework, for both the normal ordering (NO) and inverted ordering (IO) neutrino mass hierarchies. For NO (left panel), the predicted $\sum m_i$ values lie mostly below the cosmological upper bounds. The $|M_{ee}|$ values are within the reach of next-generation experiments. Hence NO is clearly allowed. For IO (right panel), the predicted $\sum m_i$ values lie mostly below the cosmological upper bounds. A part $|M_{ee}|$ values (the down line in right panel) are within the reach of next-generation experiments. Hence IO is allowed.

From Fig.1, we observe that a small but non-zero allowed
region remains. The latest oscillation data (NuFIT 6.1) slightly adjust the boundaries but
do not alter this qualitative conclusion. i.e.,  the predicted  $\sum m_i$ derived from the approximate $\mu-\tau$
reflection symmetric neutrino mass matrices $M_{23}$ for both NO and IO are allowed from the
newest data\cite{azz}.

\section{Summary and outlook}\label{sec5}

In the neutrino physics, resolving mass ordering is a primary concern. In summary,
our critical examination of the inference regarding the resolution of the neutrino mass ordering presented in Ref. [1] reveals a fundamental flaw in its underlying assumptions and methodology. Specifically, the predicted value of $\sum m_i$—derived from the approximate $\mu-\tau$ reflection-symmetric neutrino mass matrices $M_{23}$—is consistent with the latest experimental data for both NO and IO of neutrino masses.

When this condition is properly imposed, the previously claimed exclusion of inverted
mass ordering for TBM, BM, GRM, and HM is no longer valid. A small but non-zero allowed
region remains. The latest oscillation data (NuFIT 6.1) slightly adjust the boundaries but
do not alter this qualitative conclusion. 

As Singh\cite{ms} pointed, complementarity among the on-axis DUNE and off-axis T2HK experiments can enhance the sensitivity and decrease uncertainties to measure neutrino oscillation parameters. In Ref.\cite{sd}, the author projects in 2040 that the combined telescopes include IceCube, IceCube-Gen2 \cite{icg2}, Baikal-GVD\cite{gvd}, KM3NeT\cite{km3}, P-ONE\cite{po}, and TAMBO \cite{tam} and upcoming neutrinos have flavor-measuring capabilities similar to those of IceCube. Therefore, experimental determination of the neutrino mass ordering in the near future is anticipated to provide independent verification of the conclusions presented in this paper.

\section*{Acknowledgments}

This research was supported in part by the National Natural Science Foundation of China under grants No. 11875306, No. 11875062, No.11005033 and No. 12275335.

\end{document}